\documentclass[nofootinbib,aps,11pt]{revtex4-1}
\usepackage{graphicx}
\usepackage{hyperref}
\usepackage{cancel}
\usepackage{amssymb}
\usepackage{textcomp}
\usepackage{amsmath}
\usepackage{bm}
\usepackage{times}
\usepackage{epsfig}
\usepackage{color}
\begin{document}
\title{\Large Seesaw Portal to Super Heavy Dark Matter with $Z_3$ Symmetry}
\bigskip
\author{Cai-Xia Yang$^1$}
\author{Zhi-Long Han$^1$}
\email{sps\_hanzl@ujn.edu.cn}
\author{Fei Huang$^1$}
\email{sps\_huangf@ujn.edu.cn}
\author{Yi Jin$^{1,2}$}
\author{Honglei Li$^1$}
\email{sps\_lihl@ujn.edu.cn}
\affiliation{
$^1$School of Physics and Technology, University of Jinan, Jinan, Shandong 250022, China
\\
$^2$Guangxi Key Laboratory of Nuclear Physics and Nuclear Technology, Guangxi Normal University, Guilin, Guangxi 541004, China}
\date{\today}

\begin{abstract}
Right-handed neutrinos $N$ are introduced to explain the origin of the tiny neutrino masses via the seesaw mechanism. Required by relatively large Yukawa coupling and leptogenesis, masses of right-handed neutrinos are beyond $10^{9}$ GeV. Such heavy right-handed neutrino can mediate the production of super heavy dark matter $\chi$ via the freeze-in mechanism. In the minimal $Z_2$ symmetric model, the right-hand neutrino portal interaction is $y_N \phi \bar{\chi} N$ with the dark scalar $\phi$. One drawback of the $Z_2$ symmetric model is that the mass ordering $m_N>m_\phi$ with long-lived $\phi$ is almost ruled out by Big Bang Nucleosynthesis. In this paper, we propose that by extending the dark symmetry to $Z_3$, one additional interaction $y_\chi \phi \bar{\chi}^c \chi$ is further allowed. In this way, the new decay mode $\phi\to \chi\chi$ would lead to the dark scalar $\phi$ being short-lived even with a feeble $y_\chi$, thus it is allowed by the cosmological constraints. The phenomenology of the $Z_3$ symmetric super heavy dark matter model is also studied in this paper.  
\end{abstract}

\maketitle

%%%%%%%%%%%%%%%%%%%%%%%
\section{Introduction}
%%%%%%%%%%%%%%%%%%%%%%%

The neutrino oscillation experiments provide concrete evidence for new physics beyond the Standard Model (SM). One appealing pathway to explain the tiny neutrino masses is introducing the right-handed neutrinos $N$ \cite{Minkowski:1977sc,Mohapatra:1979ia,Schechter:1980gr}. To obtain sub-eV neutrino masses, the masses of right-handed neutrinos are around $10^{14}$ GeV with the Yukawa coupling $y_\nu\sim\mathcal{O}(0.1)$. Meanwhile, the success of leptogenesis to explain the origin of baryon asymmetry in the Universe also requires $m_N\gtrsim10^{9}$ GeV with hierarchical right-handed neutrino masses \cite{Davidson:2002qv}. 

Another missing piece of the Standard Model is the dark matter $\chi$. One of the most attractive candidates is the Weakly Interacting Massive Particle (WIMP). The observed relic density can be obtained with electroweak scale dark matter mass via the freeze-out mechanism \cite{Cirelli:2024ssz}. Clearly, the WIMP dark matter mass is far below the natural right-handed neutrino scale. To address the common origin of neutrino mass and dark matter, the extensively studied scenario is lowering the right-handed neutrino mass to the electroweak scale \cite{Escudero:2016tzx,Escudero:2016ksa,Tang:2016sib,Batell:2017rol,Batell:2017cmf,Bandyopadhyay:2018qcv,Ballett:2019pyw,Biswas:2021kio,Hall:2021zsk,Coito:2022kif,Liu:2022rst,Coy:2022xfj,Li:2022bpp,Liu:2024esf,Borboruah:2024lli,Abdelrahim:2025fiz}. As the WIMP dark matter is constrained by various experiments \cite{Arcadi:2017kky}, the Feebly Interacting Massive Particle (FIMP) case is also well studied \cite{Falkowski:2017uya,Becker:2018rve,Bian:2018mkl,Liu:2020mxj,Cosme:2020mck,Bandyopadhyay:2020qpn,Cheng:2020gut,Coy:2021sse,Chang:2021ose,Biswas:2022vkq,Barman:2022scg,Liu:2022cct,Herrero-Garcia:2024tyh}.

On the other hand, the super heavy dark matter near the natural right-handed neutrino scale is also viable through the freeze-in mechanism \cite{Chianese:2018dsz,Chianese:2019epo,Chianese:2020khl,Chianese:2021toe}. In the simplest $Z_2$ symmetric model, the right-handed neutrino portal interaction is $y_N\phi \bar{\chi} N$, where $\phi$ is a dark scalar. However, in the case with mass ordering $m_N>m_\phi$, the dominant decay mode of $\phi$ is into dark matter $\chi$ and SM final states via the off-shell $N$. This decay is suppressed by the feeble coupling $y_N$ and heavy right-handed neutrino mass, hence $\phi$ is usually long-lived \cite{Chianese:2018dsz}. The production of electromagnetic and hadronic showers from late decay of $\phi$ can greatly affect the Big Bang Nucleosynthesis  (BBN) and even the Cosmic Microwave Background (CMB), so the mass ordering $m_N>m_\phi$ is strongly disfavored.

In this paper, we propose the model of a seesaw portal to super heavy dark matter with $Z_3$ symmetry. Under the discrete $Z_3$ symmetry, the dark sector transforms as $\chi\to e^{i2\pi/3}\chi$ and $\phi\to e^{i2\pi/3} \phi$, which allows one additional interaction $y_\chi \phi \bar{\chi}^c\chi$ of the dark scalar $\phi$ \cite{Ghosh:2023ocl,Liu:2023kil,Liu:2023zah}. Such interaction induces the new decay mode $\phi\to \chi\chi$ of dark scalar. For feeble $y_\chi$, the decay mode $\phi\to\chi\chi$ is enough to make sure $\phi$ is short-lived when $m_\phi >2 m_\chi$, hence avoiding constraints from BBN and CMB. In this way, the mass ordering $m_N>m_\phi$ is also viable.

In Section \ref{Sec:MD}, we briefly introduce the super heavy dark matter model with $Z_3$ symmetry. The lifetime of the dark scalar is also considered in this section. In Section \ref{Sec:BE}, the Boltzmann equations of the super heavy dark sector are discussed. In Section \ref{Sec:DS}, we explore the viable parameter space and consider the impact of low reheating temperature.  Finally, conclusions are presented in Section \ref{Sec:CL}.

\section{The Model}\label{Sec:MD}

The right-handed neutrino $N$ mediates the interaction of the dark sector, which contains the dark scalar $\phi$ and the dark fermion $\chi$. In this paper, we consider $m_\phi>m_\chi$, thus the dark matter candidate is $\chi$. The relevant new interactions under the $Z_3$ symmetry are
\begin{eqnarray}\label{Eqn:Yuk}
	-\mathcal{L}&\supset&\left(y_\nu \bar{L} \tilde{H} N + y_N \phi \bar{\chi} N + \text{h.c. } \right)+ y_\chi \phi \bar{\chi}^c\chi 
	+\lambda_{H\phi} (H^\dag H) (\phi^\dag \phi) + \left(\frac{\mu}{2}\phi^3+\text{h.c. }\right),
\end{eqnarray}
where $L$ is the left-handed lepton doublet and $H$ is the SM Higgs doublet. In this paper, we focus on the Yukawa portal interactions and assume vanishing Higgs portal coupling $\lambda_{H\phi}$ and trilinear coupling $\mu$ for simplicity, which is allowed by the theoretical constraint \cite{Hektor:2019ote}. The canonical WIMP dark matter is required to be less than $\mathcal{O}(100)$ TeV by the perturbative unitarity \cite{Griest:1989wd}. In this paper, we focus on the scenario of super heavy dark matter beyond $\mathcal{O}(100)$ TeV.

After the electroweak symmetry breaking, the light neutrino masses are generated as
\begin{equation}
	m_\nu = -\frac{v_H^2}{2} y_\nu m_N^{-1} y_\nu^{T},
\end{equation}
where $v_H=246$ GeV is the vacuum expectation value of SM Higgs. Then the neutrino Yukawa coupling $y_{\nu}$ can be determined as
\begin{equation}\label{Eqn:yv}
	y_\nu=\sqrt{\frac{2 m_\nu m_N}{v_H^2}}.
\end{equation}
Typically, for $m_\nu\sim0.1$ eV, we have $y_\nu^2\sim3.3\times10^{-15} m_N/(\text{GeV})$.

\subsection{The Lifetime of Dark Scalar}
As the lifetime of the dark scalar $\phi$ plays a crucial role in the constraints from BBN, we first consider the viable decay modes of dark scalar $\phi$. The Yukawa interaction in Equation \ref{Eqn:Yuk} could directly induce the two-body decay $\phi\to N \chi$ when $m_\phi>m_N+m_\chi$ and $\phi\to \chi\chi$ when $m_\phi>2 m_\chi$. Here, the decay mode $\phi\to\chi\chi$ only appears in the $Z_3$ symmetric model. When kinematically allowed, the partial decay widths of these channels are
\begin{eqnarray}
	\Gamma(\phi\to N\chi) &=& \frac{y_N^2}{8\pi} m_\phi \left(1-\left(\frac{m_\chi+m_N}{m_\phi}\right)^2\right) \lambda^{1/2}\left(1,\frac{m_\chi^2}{m_\phi^2},\frac{m_N^2}{m_\phi^2}\right),\\
	\Gamma(\phi\to \chi \chi) &=& \frac {y_{\chi}^{2}}{4\pi}m_\phi\left(1-\frac{4m_\chi^2}{m_\phi^2}\right)^{3/2},
\end{eqnarray}
where $\lambda(x,y,z)=x^2+y^2+z^2-2xy-2xz-2yz$ is the Kallen function.

\begin{figure}
	\begin{center}
		\includegraphics[width=0.45\linewidth]{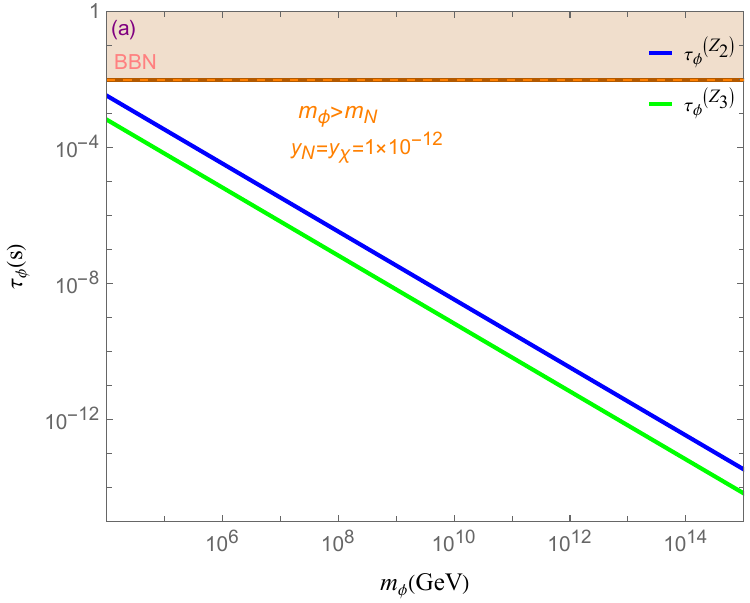}
		\includegraphics[width=0.45\linewidth]{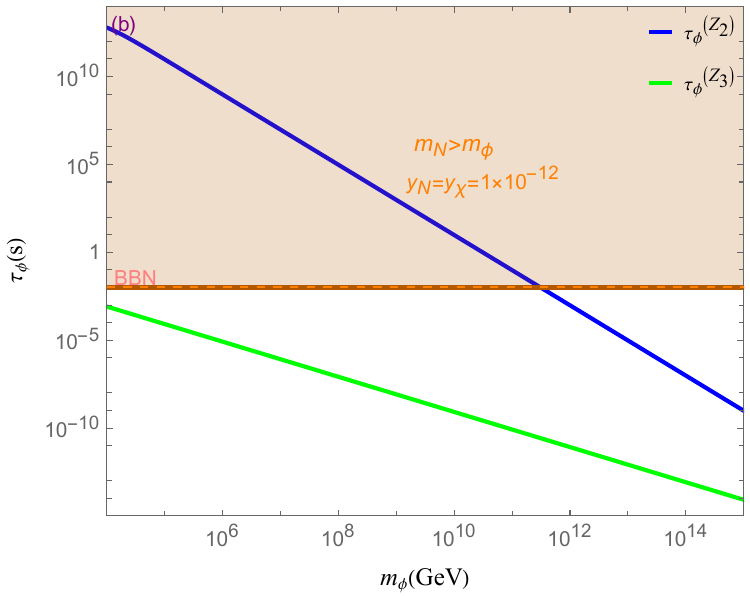}
	\end{center}
	\caption{The lifetime of dark scalar in scenario with $m_\phi>m_N$ (left) and $m_\phi<m_N$ (right). The blue and green lines are the results of $Z_2$ and $Z_3$ symmetric models, respectively. The orange region with $\tau_\phi>10^{-2}$ s is disfavored by BBN. In panel (a), we fix the mass relation as $m_\phi/m_\chi=m_\chi/m_N=10^2$. In panel (b), we fix the mass relation as $m_N/m_\phi=m_\phi/m_\chi=10^2$. }
	\label{fig1}
\end{figure}

When $m_\phi < m_N$, the dark scalar can decay into dark matter and neutrino as $\phi\to \chi \nu$ due to mixing between the light and heavy neutrino. Meanwhile, the right-handed neutrino could also mediate three-body decays of dark scalar as $\phi\to \chi h \nu$. Although  suppressed by phase space, this three-body decay might become the dominant channel for $m_\phi\gg v$. The corresponding partial decay widths in this scenario are 
\begin{eqnarray}
	\Gamma(\phi\to \chi \nu) &=& \frac {y_{N}^{2} y_\nu^2 v_H^2\, m_\phi}{16\pi m^2_N}{\left(1-\frac{m_\chi^2}{m_\phi^2}\right)^2},	\\
	\Gamma({\phi}\to\chi h\nu) & \simeq &\frac{y_{N}^{2} {y}_{\nu}^{2}}{1536 \pi^{3}} \frac{m_{\phi}^{3}}{m_{N}^{2}}\left(1+\frac{m_{\phi}^{2}}{2 m_{N}^{2}}\right),
\end{eqnarray}
where the three-body decay width is calculated in the limit $m_\phi\gg m_\chi$ \cite{Chianese:2018dsz}.

In Figure \ref{fig1}, we show the theoretical lifetime of dark scalar in the scenario with $m_\phi>m_N$ and $m_\phi<m_N$. For illustration, we have fixed the relevant Yukawa couplings to the typical freeze-in value as $y_N=y_\chi=10^{-12}$ \cite{Hall:2009bx}.
Due to the hadronic decays of SM Higgs $h$ or right-handed neutrino $N$, the lifetime of the dark scalar is constrained to be less than $10^{-2}$ s by the BBN \cite{Kawasaki:2004qu,Jedamzik:2006xz}. With additional contribution from $\phi\to \chi\chi$ in the $Z_3$ symmetric model, the lifetime of the dark scalar is smaller than it is in the $Z_2$ symmetric model. For both the $Z_2$ and $Z_3$ scenarios with $m_\phi>m_N$, it is clear in panel (a) of Figure \ref{fig1} that the lifetimes of dark scalar are always smaller than $10^{-2}$ s when $m_\phi>10^4$ GeV, thus the BBN constrain can be easily satisfied. However, in the $Z_2$ scenario with $m_\phi<m_N$, the decay width of $\phi\to \chi\nu$ is suppressed by the small light and heavy neutrino mixing, meanwhile the decay width of $\phi\to \chi h \nu$ is typically suppressed by the three-body phase space. Therefore, the dark scalar in the $Z_2$ model is long-lived. From the results in panel (b) of Figure \ref{fig1}, we find that the $Z_2$ model with $m_\phi\lesssim10^{11}$ GeV is disfavored by BBN. On the other hand, the pair decay $\phi\to \chi\chi$ is allowed in the $Z_3$ symmetric model, which makes the dark scalar short-lived even with feeble coupling for $m_\phi>10^4$ GeV. So the $Z_3$ symmetric scenario with $m_\phi<m_N$ is also allowed without fine tuning the mass spectrum \cite{Chianese:2018dsz,Bandyopadhyay:2020qpn}.

\section{Boltzmann Equations}\label{Sec:BE}

\allowdisplaybreaks

In this paper, we investigate the super heavy FIMP dark matter scenario with $y_N,y_\chi\ll 1$ under the $Z_3$ symmetry. Meanwhile, we also assume that the right-handed neutrino $N$ is also far beyond the electroweak scale. In this way,  the seesaw determined Yukawa coupling $y_\nu$ in Equation \ref{Eqn:yv} is large enough to keep the right-handed neutrino in thermal equilibrium \cite{Li:2022bpp}. Thus, the description of the abundance of dark particles requires only two coupled Boltzmann equations as

\begin{align}\label{Eqn:BE1}
    \frac{\mathrm{d} Y_{\phi}}{\mathrm{d} \mathit{x}} =&-\frac{\mathfrak{s}(x)}{\mathcal{H}(x)\mathit{x}}\langle \sigma v \rangle({\phi\phi}\rightarrow NN)\left ( Y_\phi^2 -\left ( Y^{\mathrm{eq}}_\phi \right )^2 \right )-\frac{\mathfrak{s}(x)}{\mathcal{H}(x)\mathit{x}}\langle \sigma v \rangle({\phi\chi}\rightarrow{h\nu})\left(Y_\phi Y_\chi-Y^{\mathrm{eq}}_\phi Y^{\mathrm{eq}}_\chi \right) \nonumber \\   
&-\frac{\mathfrak{s}(x)}{\mathcal{H}(x)\mathit{x}}\langle \sigma v \rangle({\phi\phi\rightarrow\chi\chi}) \left (Y_\phi^2-\frac{(Y^{\mathrm{eq}}_\phi)^2} {\left (Y^{\mathrm{eq}}_\chi\right)^2}  Y_\chi^2 \right )
- \frac{\mathfrak{s}(x)}{\mathcal{H}(x)\mathit{x}}\langle \sigma v \rangle({\phi\phi}\rightarrow N\chi)\left ( Y_\phi^2 -\frac{ (Y^{\mathrm{eq}}_\phi)^2}{Y^{\mathrm{eq}}_{\chi} } Y_\chi \right ) \nonumber\\
 &+\frac{ \tilde{\Gamma}({N\rightarrow \phi\chi}) }{\mathcal{H}(x)\mathit{x}}\left ( Y_N^{\mathrm{eq}}-\frac{Y_N^{\mathrm{eq}}}{Y^{\mathrm{eq}}_\phi Y^{\mathrm{eq}}_\chi}Y_\phi Y_\chi \right )-\frac{\tilde{\Gamma}({\phi\rightarrow\chi\chi})}{\mathcal{H}(x)\mathit{x}} \left (Y_\phi-\frac{Y^{\mathrm{eq}}_\phi} {\left (Y^{\mathrm{eq}}_\chi\right)^2}  Y_\chi^2 \right )\nonumber\\
&-\frac{\tilde{\Gamma}(\phi\rightarrow \chi N,\chi \nu,\chi h \nu)}{\mathcal{H}(x)\mathit{x}} \left (Y_\phi-\frac{Y^{\mathrm{eq}}_\phi} {Y^{\mathrm{eq}}_\chi}  Y_\chi \right )
\end{align}

\begin{align}\label{Eqn:BE2}
    \frac{\mathrm{d} Y_{\chi}}{\mathrm{d} \mathit{x}} =&-\frac{\mathfrak{s}(x)}{\mathcal{H}(x)\mathit{x}}\langle \sigma v \rangle({\chi\chi}\rightarrow NN)\left ( Y_\chi^2- \left ( Y^{\mathrm{eq}}_\chi \right )^2 \right )-\frac{\mathfrak{s}(x)}{\mathcal{H}(x)\mathit{x}}\langle \sigma v \rangle({\phi\chi}\rightarrow{h\nu}) \left(Y_\phi Y_\chi-Y^{\mathrm{eq}}_\phi Y^{\mathrm{eq}}_\chi \right) \nonumber\\    
    &+\frac{\mathfrak{s}(x)}{\mathcal{H}(x)\mathit{x}}\langle \sigma v \rangle({\phi\phi\rightarrow\chi\chi}) \left (Y_\phi^2-\frac{(Y^{\mathrm{eq}}_\phi)^2} {\left (Y^{\mathrm{eq}}_\chi\right)^2}  Y_\chi^2 \right )
    + \frac{\mathfrak{s}(x)}{\mathcal{H}(x)\mathit{x}}\langle \sigma v \rangle({\phi\phi}\rightarrow N\chi)\left ( Y_\phi^2 -\frac{ (Y^{\mathrm{eq}}_\phi)^2}{Y^{\mathrm{eq}}_{\chi} } Y_\chi \right )  \nonumber\\
    &- \frac{\mathfrak{s}(x)}{\mathcal{H}(x)\mathit{x}}\langle \sigma v \rangle({\chi\chi}\rightarrow N\chi)\left ( Y_\chi^2- Y^{\mathrm{eq}}_\chi Y_\chi\right )
    -\frac{\mathfrak{s}(x)}{\mathcal{H}(x)\mathit{x}}\langle \sigma v \rangle({\phi\chi}\rightarrow N\phi)\left (Y_\phi Y_{\chi}- Y^{\mathrm{eq}}_\chi Y_\phi \right ) \nonumber\\
    &+\frac{ \tilde{\Gamma}({N\rightarrow \phi\chi}) }{\mathcal{H}(x)\mathit{x}}\left ( Y_N^{\mathrm{eq}}-\frac{Y_N^{\mathrm{eq}}}{Y^{\mathrm{eq}}_\phi Y^{\mathrm{eq}}_\chi}Y_\phi Y_\chi \right )+2\frac{\tilde{\Gamma}({\phi\rightarrow\chi\chi})}{\mathcal{H}(x)\mathit{x}} \left (Y_\phi-\frac{Y^{\mathrm{eq}}_\phi} {\left (Y^{\mathrm{eq}}_\chi\right)^2}  Y_\chi^2 \right ) \nonumber\\
    &+\frac{\tilde{\Gamma}({\phi\rightarrow \chi N,\chi \nu,\chi h \nu})}{\mathcal{H}(x)\mathit{x}} \left (Y_\phi-\frac{Y^{\mathrm{eq}}_\phi} {Y^{\mathrm{eq}}_\chi}  Y_\chi \right )
\end{align}
where $ x=m_N/T $. The abundance is defined as $Y_i=n_i/\mathfrak{s}$, where $n_i$ is the number density of particle $i$. The entropy density $\mathfrak{s}$ and Hubble parameter $\mathcal{H}$ are
\begin{equation}
	\mathfrak{s}(x)=\frac{2\pi^2}{45} g_{*} \frac{m_N^3}{x^3},\quad 
	\mathcal{H}(x) =\sqrt{\frac{4\pi^3 g_{*}}{45}} \frac{m_N^2}{M_\text{pl} x^2},
\end{equation}
where $g_{*}$ is the effective number of relativistic degrees of freedom, and $M_\text{pl}$ is the Planck mass. $\langle \sigma v \rangle$ is the thermal averaged cross section, which is calculated numerically by micrOMEGAs \cite{Belanger:2013oya}. The thermal decay width is denoted as
\begin{equation}
	\tilde{\Gamma}(A\to B C)= \Gamma(A\to B C) \frac{\mathcal{K}_1(m_A/T)}{\mathcal{K}_2(m_A/T)},
\end{equation}
where $\mathcal{K}_1$ and $\mathcal{K}_2$ are the first and second modified Bessel functions of the second kind. The dark decay width of right-handed neutrino $N\to\phi\chi$ is
\begin{equation}
	\Gamma(N\to \phi\chi) =\frac{y_N^2}{16\pi} m_N \left(\left(1+\frac{m_\chi}{m_N}\right)^2-\frac{m_\phi^2}{m_N^2}\right) \lambda^{1/2}\left(\frac{m_\phi^2}{m_N^2},\frac{m_\chi^2}{m_N^2},1\right).
\end{equation}

The Equations \eqref{Eqn:BE1} and \eqref{Eqn:BE2} contain the viable production channels of dark particles in the $Z_3$ symmetric model. For the dark sector scattering processes $ \phi\phi,\chi\chi\to NN$, their annihilation cross sections are proportional to the coupling of $y_N^4$. Meanwhile, the cross section of the neutrino Yukawa scattering process $\phi\chi\to h\nu$ is proportional to $y_N^2 y_\nu^2$, where $y_\nu$ is determined by the seesaw relation in Equation \ref{Eqn:yv}. For the right-handed neutrino well the above electroweak scale, we usually have $y_\nu\simeq5.7\times10^{-8}\times\sqrt{m_N/(\text{GeV})}\gg\mathcal{O}(10^{-7})\gg y_N$. So the contribution from the neutrino Yukawa scattering process $\phi\chi\to h \nu$ is usually larger than the contribution from the dark sector scattering processes $ \phi\phi,\chi\chi\to NN$ for super heavy right-handed neutrino \cite{Chianese:2018dsz}.

It should be noted that the scattering process as $\phi\chi\to h\nu$ involves the right-handed neutrino in the $s$-channel. We then adopt the real intermediate states (RIS) subtraction strategy to avoid the double counting of resonant on-shell generation of right-handed neutrino \cite{Kolb:1979qa}. The previous analysis of super heavy dark matter does not have such double counting problem \cite{Chianese:2018dsz,Chianese:2019epo,Chianese:2020khl,Chianese:2021toe}, as they only focus on the mass spectrum $m_\phi>m_N$. In this paper, we also consider the mass spectrum $m_N>m_\phi$. The on-shell direct decay $N\to \phi \chi$ is possible in this scenario, so the RIS subtraction should be applied. 

For the sake of completeness, we also consider the possible conversion processes as $\phi\phi\to\chi\chi$,$\phi\phi\to N\phi$, $\chi\chi\to N\chi$, and $\phi\chi\to N\phi$  in the Boltzmann equations. With the same order of the dark sector Yukawa couplings $y_N\sim y_\chi$, the cross sections of these conversion processes are in the same order of the dark scattering processes $\phi\phi,\chi\chi\to NN$. In the FIMP scenario, the abundances of dark particles never reach the thermal equilibrium, i.e., $Y_{\phi,\chi}\ll Y_{\phi,\chi}^\text{eq}$, which indicates that the contributions from these conversion processes can be neglected in the numerical calculations. Therefore, the new contributions of conversion processes $\phi\phi\to N\phi$, $\chi\chi\to N\chi$, and $\phi\chi\to N\phi$ in the $Z_3$ symmetric model would not have a large impact on the evolution of dark particle abundances. On the other hand, the new decay mode $\phi\to \chi\chi$ could affect the conversion of dark scalar $\phi$ into dark matter $\chi$.

In this paper, the masses of the dark particles are much larger than the electroweak scale. They are produced via the non-thermally freeze-in mechanism with feeble couplings, which are sensitive to the initial conditions. The abundances of dark particles $\phi$ and $\chi$ are then determined by solving the Boltzmann equations with $Y_\phi=Y_\chi=0$ at the end of inflation.  These initial conditions are applied at the reheating temperature of the Universe $T_\text{RH}$, from which the Universe begins to be dominated by the radiation. The contribution of certain terms  in the Boltzmann equations to the abundance can be obtained by solving the  integration as
\begin{eqnarray}\label{Eqn:RHS}
	Y(x)\simeq\int_{x_\text{RH}}^x dx \frac{\mathfrak{s}(x)}{\mathcal{H}(x)\mathit{x}} \langle \sigma v\rangle (AB\to CD) Y_A^\text{eq} Y_B^\text{eq},	
\end{eqnarray}
for the scattering process, and
\begin{eqnarray}\label{Eqn:RHD}
	Y(x)\simeq\int_{x_\text{RH}}^x dx \frac{1}{\mathcal{H}(x)\mathit{x}} \tilde{\Gamma}(A\to BC) Y_A^\text{eq},	
\end{eqnarray}
for the decay process, where $x_\text{RH}=m_N/T_{RH}$. In the limit $m_N\ll T_\text{RH}$, $x_\text{RH}\to0$ will reproduce the result of the canonical high reheating temperature scenario. In this paper, we also consider the impact of the low reheating temperature scenario, i.e., $T_\text{RH} < m_{N}$ or $m_\phi$. The Planck \cite{Planck:2018jri} and BICEP-Keck \cite{BICEP2:2018kqh} experiments have imposed the constraint on the tensor-to-scalar ratio $r<0.056$ at the 95\% CL, which implies an upper limit on the reheating temperature 
\begin{equation}
	T_\text{RH}\leq T_\text{RH}^\text{max}\simeq6.5\times10^{15}~\text{GeV}.
\end{equation}

The present dark matter relic density is calculated as
\begin{equation}\label{Eq:Omg}
	\Omega_{\mathrm{DM}} h^{2}=\frac{2 \mathfrak{s}_{0} m_{\chi} Y_{\mathrm{DM}}^0}{\rho_{\text {crit }} / h^{2}},
\end{equation}
where $\mathfrak{s}_{0}=2891.2~\text{cm}^{-3}$ is today's entropy density, and $\rho_{\text {crit }} / h^{2}=1.054\times10^{-5}~\text{GeV}~\text{cm}^{-3}$ is the critical density \cite{ParticleDataGroup:2024cfk}. The observed dark matter relic density is $\Omega_{\mathrm{DM}} h^2 =0.120\pm0.001$ by the Planck corporation \cite{Planck:2018vyg}. The present dark matter abundance is derived as 
\begin{eqnarray}\label{Eq:YDM}
	Y_\text{DM}^0 &= &Y_\chi(\infty)+Y_\phi(\infty)\times\left[\text{BR}(\phi\to N\chi, \chi\nu,\chi h\nu) + 2~\text{BR}(\phi\to \chi\chi)\right] \nonumber \\ 
	&= &Y_\chi(\infty)+Y_\phi(\infty) \times\left[1+\text{BR}(\phi\to \chi\chi)\right],
\end{eqnarray}
where we have applied the relation $\text{BR}(\phi\to N\chi, \chi\nu,\chi h\nu) =1-\text{BR}(\phi\to \chi\chi)$. Therefore, with a new contribution from the decay mode $\phi\to \chi\chi$, the abundance of dark matter  in the $Z_3$ symmetric model is expected to be larger than it in the $Z_2$ symmetric model.

Before ending this section, we briefly comment on the contribution of the Higgs portal interaction $\lambda_{H\phi} (H^\dag H) (\phi^\dag \phi)$. With sizable Higgs coupling $\lambda_{H\phi}$ \cite{Chianese:2021toe}, the dark scalar $\phi$ might dominantly produce through the Higgs portal interactions with SM particles as $\phi\phi\to W^+W^-,ZZ,hh$ \cite{Cheng:2020gut}, which is independent of the seesaw portal coupling $y_N$. After the production of dark scalar $\phi$,  the dark matter $\chi$ is then generated via the scalar decay $\phi\to N\chi, \chi\chi$.

\section{Results and Discussion}\label{Sec:DS}

In this section, we illustrate the numerical results of the $Z_3$ symmetric model, which is determined by the free parameter set as $\{m_\chi,m_\phi, m_N, y_N, y_\chi\}$. As the decay mode $\phi\to\chi\chi$ is important to satisfy the BBN constraint, we will assume a hierarchical mass spectrum as $m_\phi\gg m_\chi\gg m_N$ or $m_N\gg m_\phi \gg m_\chi$, thus $\phi\to \chi\chi$ is always allowed in the following discussion. Special compressed mass spectrums, such as $m_\phi<2m_\chi$, $m_\phi <m_N+m_\chi$, or $m_N<m_\phi+m_\chi$ are theoretically allowed but will not be considered in this paper.

\subsection{Ordering Type A: $m_\phi>m_N, m_\chi$}\label{SEC:OTA}

We first consider the mass ordering type A with $m_\phi>m_N, m_\chi$. Neglecting certain scattering terms suppressed by the condition $Y_i\ll Y_i^\text{eq}$, the Boltzmann equations \eqref{Eqn:BE1} and \eqref{Eqn:BE2} in this scenario can be further simplified as
\begin{eqnarray}
	\frac{dY_\phi}{dx}& = &+\frac{\mathfrak{s}(x)}{\mathcal{H}(x)\mathit{x}}\langle \sigma v \rangle({\phi\phi}\rightarrow NN)\left ( Y^{\mathrm{eq}}_\phi \right )^2 +\frac{\mathfrak{s}(x)}{\mathcal{H}(x)\mathit{x}}\langle \sigma v \rangle({\phi\chi}\rightarrow{h\nu})Y^{\mathrm{eq}}_\phi Y^{\mathrm{eq}}_\chi \nonumber \\
	& & -\frac{\tilde{\Gamma}({\phi\rightarrow\chi\chi})}{\mathcal{H}(x)\mathit{x}} \left (Y_\phi-\frac{Y^{\mathrm{eq}}_\phi} {\left (Y^{\mathrm{eq}}_\chi\right)^2}  Y_\chi^2 \right ) -\frac{\tilde{\Gamma}(\phi\rightarrow \chi N)}{\mathcal{H}(x)\mathit{x}} \left (Y_\phi-\frac{Y^{\mathrm{eq}}_\phi} {Y^{\mathrm{eq}}_\chi}  Y_\chi \right ) ,\\
	\frac{dY_\chi}{dx}& = & +\frac{\mathfrak{s}(x)}{\mathcal{H}(x)\mathit{x}}\langle \sigma v \rangle({\chi\chi}\rightarrow NN) \left ( Y^{\mathrm{eq}}_\chi \right )^2 +\frac{\mathfrak{s}(x)}{\mathcal{H}(x)\mathit{x}}\langle \sigma v \rangle({\phi\chi}\rightarrow{h\nu}) Y^{\mathrm{eq}}_\phi Y^{\mathrm{eq}}_\chi \nonumber \\
	&&+2\frac{\tilde{\Gamma}({\phi\rightarrow\chi\chi})}{\mathcal{H}(x)\mathit{x}} \left (Y_\phi-\frac{Y^{\mathrm{eq}}_\phi} {\left (Y^{\mathrm{eq}}_\chi\right)^2}  Y_\chi^2 \right ) +\frac{\tilde{\Gamma}(\phi\rightarrow \chi N)}{\mathcal{H}(x)\mathit{x}} \left (Y_\phi-\frac{Y^{\mathrm{eq}}_\phi} {Y^{\mathrm{eq}}_\chi}  Y_\chi \right ) .
\end{eqnarray}
Compared with the direct two-body decays $\phi\to \chi \chi$ and $\phi\to \chi N$, the decay $\phi\to \chi\nu$ is further suppressed by the mixing between the light and heavy neutrino $\theta\sim\sqrt{m_\nu/m_N}\ll1$, so the contribution of $\phi\to\chi\nu$ is also neglected in the above simplified Boltzmann equations. Meanwhile, the result for the $Z_2$ scenario is obtained by removing the contribution from decay $\phi\to \chi\chi$.

\begin{figure}
	\begin{center}
		\includegraphics[width=0.45\linewidth]{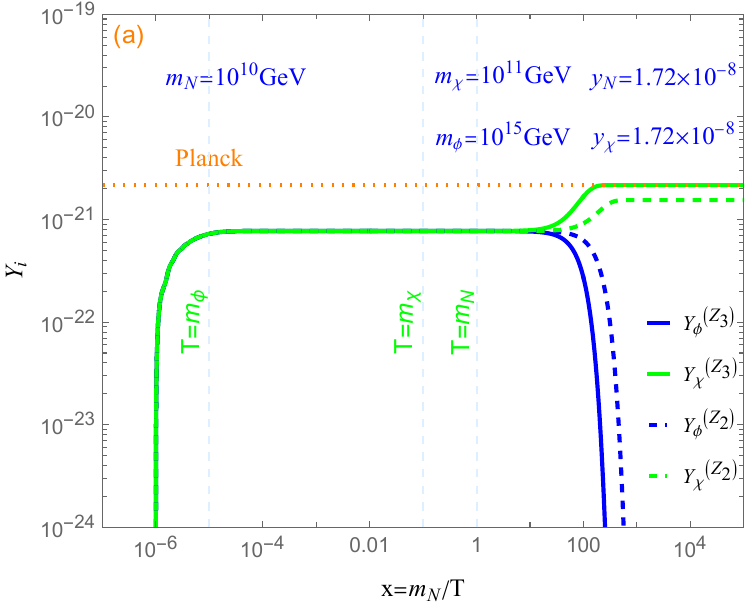}
		\includegraphics[width=0.45\linewidth]{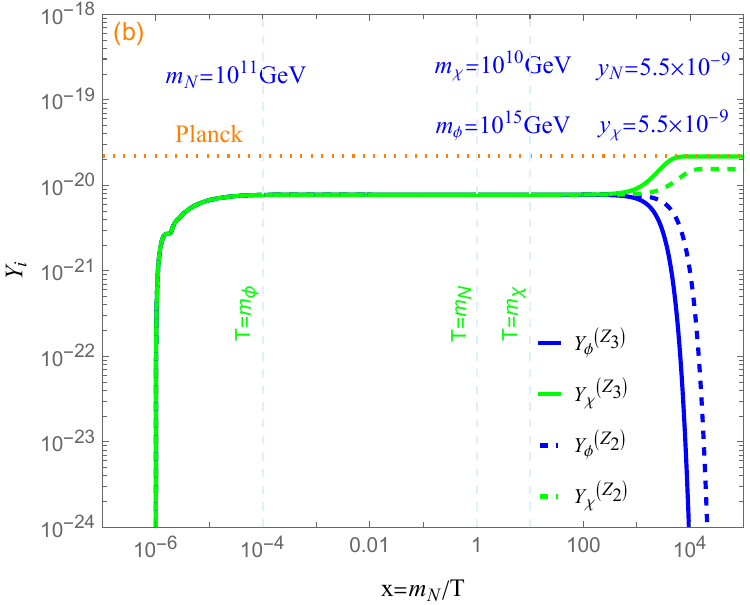}
	\end{center}
	\caption{Evolution of dark sector for ordering type A.  The solid (dashed) blue and green lines are the results for dark scalar $\phi$ and dark matter $\chi$ in the $Z_3$ ($Z_2$) symmetric model. The orange dotted lines are the required abundance of dark matter to satisfy the Planck observed result. We set $m_N=10^{10}$ GeV with $m_\chi=10^{11}$ GeV in panel (a) and $m_N=10^{11}$ GeV with  $m_\chi=10^{10}$ GeV in panel (b), meanwhile,  $m_\phi=10^{15}$ GeV is fixed. }
	\label{fig2}
\end{figure}

If we consider the right-handed neutrino mass above $10^9$ GeV for successful leptogenesis with hierarchy right-handed neutrinos \cite{Davidson:2002qv}, the contributions from the dark scattering processes $\phi\phi\to NN$ and $\chi\chi\to NN$ can be further neglected, i.e., only the neutrino Yukawa scattering process $\phi\chi\to h\nu$ is the dominant contribution \cite{Chianese:2018dsz}. In the heavy dark scalar limit $m_\phi\gg m_N, m_\chi$, the abundances of dark particles can be approximately derived as \cite{Chianese:2021toe}
\begin{eqnarray}
	Y_\chi(\infty)=Y_\phi(\infty)\approx \int_{0}^{\infty} dx \frac{\mathfrak{s}(x)}{\mathcal{H}(x)\mathit{x}} \langle \sigma v\rangle ( \phi\chi\to h\nu) Y_\phi^\text{eq} Y_\chi^\text{eq}\approx \frac{405 M_\text{pl}}{1.66 g_{*}^{3/2}2^{14} \pi^5} \frac{y_\nu^2 y_N^2}{m_\phi},
\end{eqnarray}
where $T_\text{RH}\gg m_\phi$ is also assumed. Using Equation \eqref{Eq:Omg} and \eqref{Eq:YDM}, the present dark matter abundance can be calculated as
\begin{equation}\label{Eqn:RDA}
	\Omega_{\mathrm{DM}}h^2\simeq 2.9\times10^{20}\times y_\nu^2 y_N^2 ~\frac{m_\chi}{m_\phi} ~[2+\text{BR}(\phi\to\chi\chi)].
\end{equation}
To generate the observed dark matter abundance, we typically require the Yukawa couplings $y_\nu^2 y_N^2\sim 2\times10^{-22} m_\phi/m_\chi$. Thus the dark matter abundance is sensitive to the mass ratio $m_\phi/m_\chi$. It is also worth mentioning that the contribution of the new decay channel $\phi\to\chi\chi$ only appears as the additional factor of BR$(\phi\to\chi\chi)$ in Equation \eqref{Eqn:RDA}. So the dark matter abundance in the $Z_3$ model is at most $3/2$ times as large as it is in the $Z_2$ model.

In Figure \ref{fig2}, we show the numerical results of ordering type A for two benchmark points. The panel (a) of Figure \ref{fig2} represents the scenario with $m_\phi>m_\chi>m_N$, meanwhile, the panel (b) of Figure \ref{fig2} represents the scenario with $m_\phi>m_N>m_\chi$. We also fix $y_N=y_\chi$ for the comparison of the $Z_2$ and $Z_3$ symmetry. Both scenarios correspond to the high mass right-handed neutrino for thermal leptogenesis, thus the dark matter abundance for the two benchmark points are determined by the neutrino Yukawa scattering $\phi\chi\to h \nu$. In this way, the value of $Y_\phi^\text{eq}$ is exponentially suppressed when the temperature $T\lesssim m_\phi$, which leads to the abundances of dark particles freeze-in at $T\sim m_\phi$. Later, the dark scalar $\phi$ is converted into dark matter $\chi$ via decays when $\tilde{\Gamma}(\phi)>\mathcal{H}$. In the $Z_3$ model, the additional $\phi\to\chi\chi$ decay mode makes the total decay width of $\phi$ larger than it is in the $Z_2$ model, so the conversions in the $Z_3$ model happen always earlier than in the $Z_2$ model. It is also clear that the present dark matter abundance in the $Z_3$ model is larger than it in the $Z_2$ model, due to more efficiency of $\phi\to \chi\chi$ than $\phi\to \chi N$.

\begin{figure}
	\begin{center}
		\includegraphics[width=0.45\linewidth]{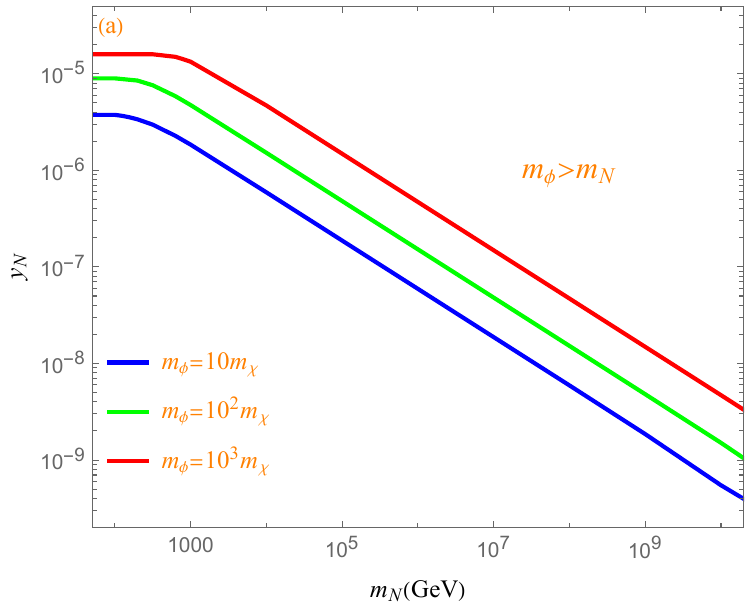}
		\includegraphics[width=0.47\linewidth]{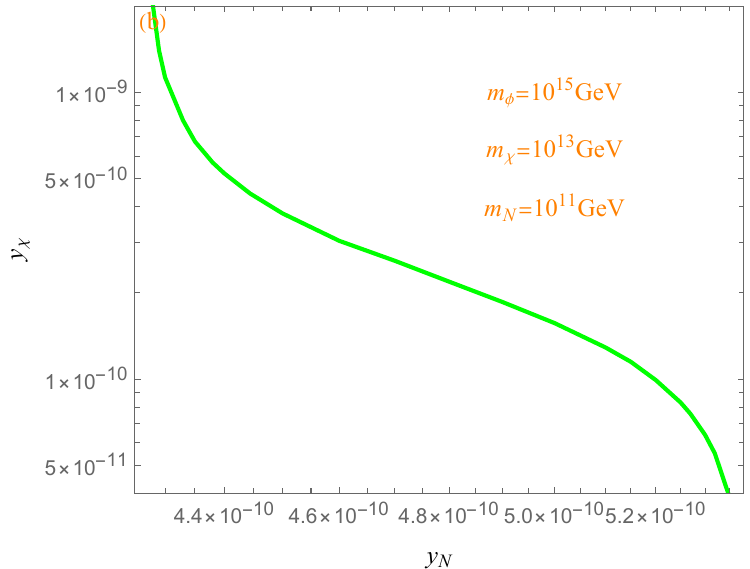}
	\end{center}
	\caption{The dependence of Yukawa couplings for ordering type A. Panel (a): The required values of the Yukawa coupling $y_N$ as a function of right-handed neutrino mass $m_N$. We also fix $y_N=y_\chi$.  Panel (b): The required values of the Yukawa coupling $y_\chi$ as a function of the Yukawa coupling $y_N$. The mass spectrum is set as $m_\phi=10^{15}$ GeV, $m_N=10^{11}$ GeV, and $m_\chi=10^{13}$ GeV. }
	\label{fig3}
\end{figure}

The dark matter abundance in Equation \eqref{Eqn:RDA} is also closely related to the right-handed neutrino mass when the neutrino Yukawa coupling is determined by the seesaw relation. Inserting Equation \eqref{Eqn:yv}, we simply expect $y_N^2 \propto 1/m_N$ for correct relic abundance when $\phi\chi\to h \nu$ is the dominant channel. In panel (a) of Figure \ref{fig3}, we show the required values of the Yukawa coupling $y_N$ as a function of right-handed neutrino mass $m_N$ for correct dark matter relic abundance. It confirms the fact that for a fixed value of mass ratio $m_\phi/m_\chi$, a larger $m_N$ also leads to a larger $y_\nu$, thus a smaller $y_N$ is required. Meanwhile, increasing the mass ratio $m_\phi/m_\chi$ results in larger $y_N$ for correct relic abundance.  On the other hand, when the right-handed neutrino is lighter than $10^4$ GeV, the seesaw induced neutrino Yukawa coupling $y_\nu\lesssim\mathcal{O}(10^{-6})$ is smaller than coupling $y_N$. So the dark scattering processes $\phi\phi,\chi\chi\to NN$ become the dominant contribution, which does not depend on the right-handed neutrino mass. In panel (b) of Figure \ref{fig3}, we then illustrate the impact of the new Yukawa coupling $y_\chi$ in the $Z_3$ model. Qualitatively speaking, the branching ratio of $\phi\to\chi\chi$ becomes larger with increasing $y_\chi$, thus a smaller $y_N$ is enough to generate correct relic abundance. Such relation is clear when $y_\chi\sim y_N$. In the limit $y_\chi\gg y_N$, the branching ratio of $\phi\to \chi\chi$ approaches to one, hence $y_N$ has the minimum value for the benchmark point.

\subsection{Ordering Type B: $m_N>m_\phi> m_\chi$}\label{SEC:OTB}

We then consider the mass ordering type B with $m_N>m_\phi>m_\chi$.  The complete Boltzmann equations \eqref{Eqn:BE1} and \eqref{Eqn:BE2} can also be simplified by neglecting certain scattering terms suppressed by the condition $Y_i\ll Y_i^\text{eq}$ as 
\begin{eqnarray}
	\frac{dY_\phi}{dx}& = &+\frac{\mathfrak{s}(x)}{\mathcal{H}(x)\mathit{x}}\langle \sigma v \rangle(NN\rightarrow {\phi\phi})\left ( Y^{\mathrm{eq}}_N \right )^2 +\frac{\mathfrak{s}(x)}{\mathcal{H}(x)\mathit{x}}\langle \sigma v \rangle^\text{sub}({\phi\chi}\rightarrow{h\nu})Y^{\mathrm{eq}}_\phi Y^{\mathrm{eq}}_\chi \nonumber \\
	& & +\frac{ \tilde{\Gamma}({N\rightarrow \phi\chi}) }{\mathcal{H}(x)\mathit{x}}\left ( Y_N^{\mathrm{eq}}-\frac{Y_N^{\mathrm{eq}}}{Y^{\mathrm{eq}}_\phi Y^{\mathrm{eq}}_\chi}Y_\phi Y_\chi \right ) -\frac{\tilde{\Gamma}({\phi\rightarrow\chi\chi})}{\mathcal{H}(x)\mathit{x}} \left (Y_\phi-\frac{Y^{\mathrm{eq}}_\phi} {\left (Y^{\mathrm{eq}}_\chi\right)^2}  Y_\chi^2 \right )\\ \nonumber 
	&&-\frac{\tilde{\Gamma}({\phi\rightarrow \chi \nu,\chi h \nu})}{\mathcal{H}(x)\mathit{x}} \left (Y_\phi-\frac{Y^{\mathrm{eq}}_\phi} {Y^{\mathrm{eq}}_\chi}  Y_\chi \right ) \\
	\frac{dY_\chi}{dx}& = & +\frac{\mathfrak{s}(x)}{\mathcal{H}(x)\mathit{x}}\langle \sigma v \rangle( NN\rightarrow {\chi\chi}) \left ( Y^{\mathrm{eq}}_N \right )^2 +\frac{\mathfrak{s}(x)}{\mathcal{H}(x)\mathit{x}}\langle \sigma v \rangle^\text{sub}({\phi\chi}\rightarrow{h\nu}) Y^{\mathrm{eq}}_\phi Y^{\mathrm{eq}}_\chi \nonumber \\
	&&+\frac{ \tilde{\Gamma}({N\rightarrow \phi\chi}) }{\mathcal{H}(x)\mathit{x}}\left ( Y_N^{\mathrm{eq}}-\frac{Y_N^{\mathrm{eq}}}{Y^{\mathrm{eq}}_\phi Y^{\mathrm{eq}}_\chi}Y_\phi Y_\chi \right) +2\frac{\tilde{\Gamma}({\phi\rightarrow\chi\chi})}{\mathcal{H}(x)\mathit{x}} \left (Y_\phi-\frac{Y^{\mathrm{eq}}_\phi} {\left (Y^{\mathrm{eq}}_\chi\right)^2}  Y_\chi^2 \right ) \\ \nonumber
	&&+\frac{\tilde{\Gamma}({\phi\rightarrow \chi \nu,\chi h \nu})}{\mathcal{H}(x)\mathit{x}} \left (Y_\phi-\frac{Y^{\mathrm{eq}}_\phi} {Y^{\mathrm{eq}}_\chi}  Y_\chi \right ),
\end{eqnarray}
where $\langle \sigma v \rangle^\text{sub}({\phi\chi}\rightarrow{h\nu})$ is the RIS subtracted annihilation cross section \cite{Kolb:1979qa}. In this scenario, the two-body decay $N\to \phi\chi$ is the dominant contribution in the high reheating temperature limit $T_\text{RH}\gg m_N$, the contributions from the scattering terms as $NN\to \phi\phi,\chi\chi$ and $\phi\chi\to h\nu$ can be neglected in the numerical calculation \cite{Chianese:2018dsz}. With the same order of Yukawa couplings $y_\chi\simeq y_N$, the decay width $\phi\to \chi\chi$ is also much larger than the decay widths $\phi\to \chi\nu, \chi h\nu$. Therefore, we  can only consider the $\phi\to\chi\chi$ term in the $Z_3$ model. On the other hand, $\phi\to\chi\nu,\chi h\nu$ should be considered in the $Z_2$ model due to the absence of $\phi\to \chi\chi$.

Provided $T_\text{RH}\gg m_N$, the final abundances of dark particles from $N\to \phi\chi$ decay is calculated as
\begin{equation}
	Y_\chi(\infty)=Y_\phi(\infty)\approx \int_{0}^{\infty}dx \frac{1}{\mathcal{H}(x) x} \tilde{\Gamma}(N\to\phi\chi) Y_N^\text{eq}\approx \frac{135 M_\text{pl}}{1.66 g_{*}^{3/2} 2 \pi^3} \frac{\Gamma(N\to \phi\chi)}{m_N^2}.
\end{equation}
So the present dark matter abundance is
\begin{equation}\label{Eqn:RDB}
	\Omega_{\mathrm{DM}}h^2 \simeq 1.6\times10^{23} \times y_N^2 \frac{m_\chi}{m_N}[2+\text{BR}(\phi\to\chi\chi)].
\end{equation}
Correct dark matter abundance can be obtained with $y_N^2\sim4\times10^{-25} m_N/m_\chi$. Compared with the scattering process $\phi\chi\to h\nu$, the two-body decay $N\to \phi\chi$ requires much smaller Yukawa coupling $y_N$. It also indicates that the dark matter abundance is sensitive to the mass ratio $m_N/m_\chi$ rather than the right-handed neutrino mass itself.

\begin{figure}
	\begin{center}
		\includegraphics[width=0.45\linewidth]{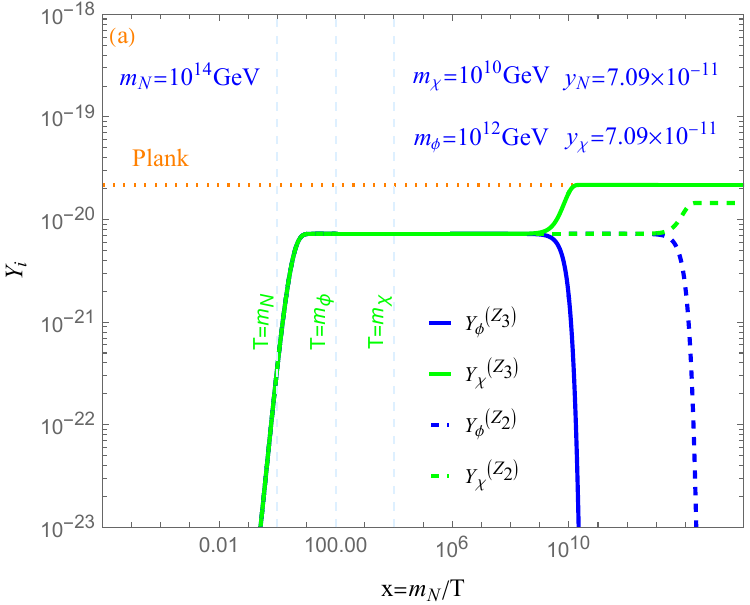}
		\includegraphics[width=0.45\linewidth]{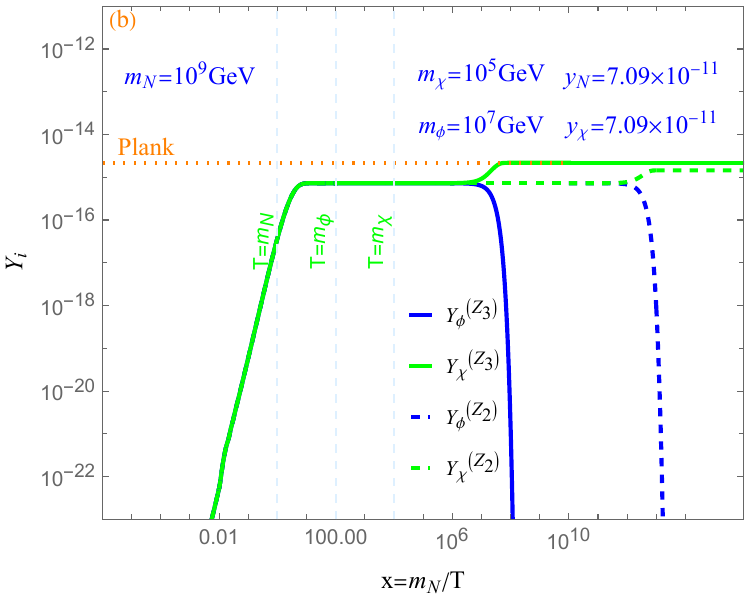}
	\end{center}
	\caption{Evolution of dark sector for ordering type B.  The solid (dashed) blue and green lines are the results for dark scalar $\phi$ and dark matter $\chi$ in the $Z_3$ ($Z_2$) symmetric model. The orange dotted lines are the required abundance of dark matter to satisfy the Planck observed result. We set $m_N=10^{14}$ GeV, $m_\phi=10^{12}$ GeV, $m_\chi=10^{10}$ GeV in panel (a), and $m_N=10^{9}$, $m_\phi=10^{7}$ GeV, $m_\chi=10^{5}$ GeV in panel (b). }
	\label{fig4}
\end{figure}

In Figure \ref{fig4}, we show the numerical results of ordering type B for two benchmark points. In panel (a) of Figure \ref{fig4}, we show a benchmark point in $Z_2$ model with $\phi$ decaying before BBN, meanwhile, a benchmark point in $Z_2$ model with $\phi$ decaying after BBN is in panel (b). For comparison, we also set $y_N=y_\chi$ in the $Z_3$ model. The values of the Yukawa couplings for the benchmark points are obtained by requiring the correct relic density of the $Z_3$ model. Meanwhile, the final dark matter abundance in the $Z_2$ model is always smaller than it is in the $Z_3$ model.

As the dark particles are dominantly generated via the decay $N\to\phi\chi$, their abundances are freeze-in at the temperature of $T\sim m_N$. Then, the conversion of dark scalar $\phi$ into dark matter $\chi$ happens at $\tilde{\Gamma}(\phi)>\mathcal{H}$. In the $Z_3$ model, the direct two-body decay $\phi\to\chi\chi$ leads to the conversion of $\phi\to\chi$ at the temperature of $T\sim\mathcal{O}(10^2 -10^4)$ GeV for the benchmark points. In contrast, the conversion of $\phi\to\chi$ in the $Z_2$ model is dominant by the three-body decay $\phi\to \chi h\nu$ for dark scalar mass above the TeV scale. As $\Gamma(\phi\to \chi h\nu)\ll \Gamma(\phi\to \chi\chi)$ with $y_N=y_\chi$, the conversion of dark scalar in the $Z_2$ model happens much later than in the $Z_3$ model. For the two benchmark points in Figure \ref{fig4}, we typically have $\phi$ decays in the $Z_2$ model at the temperature of $T\sim1$ GeV for panel (a) and $T\sim0.1$ MeV $< T_\text{BBN}\sim 4$ MeV \cite{Sarkar:1995dd} for panel (b). Therefore, the latter case is excluded by BBN.

\begin{figure}
	\begin{center}
		\includegraphics[width=0.45\linewidth]{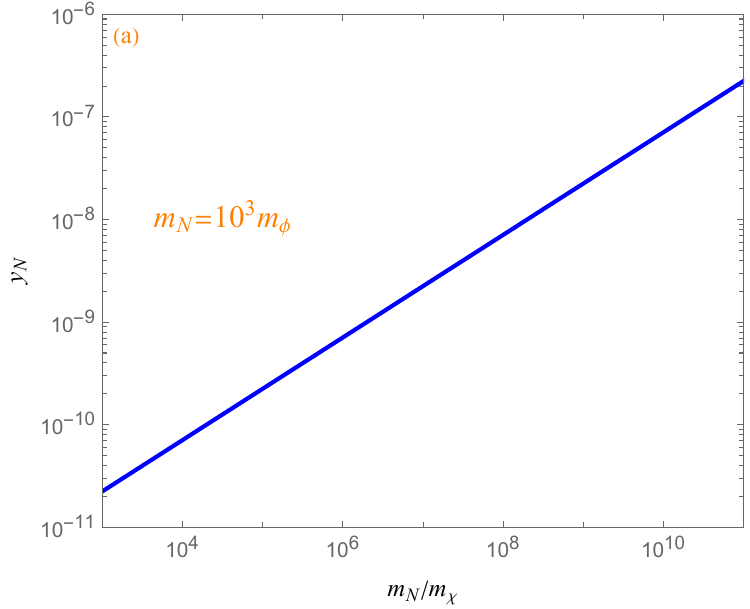}
		\includegraphics[width=0.46\linewidth]{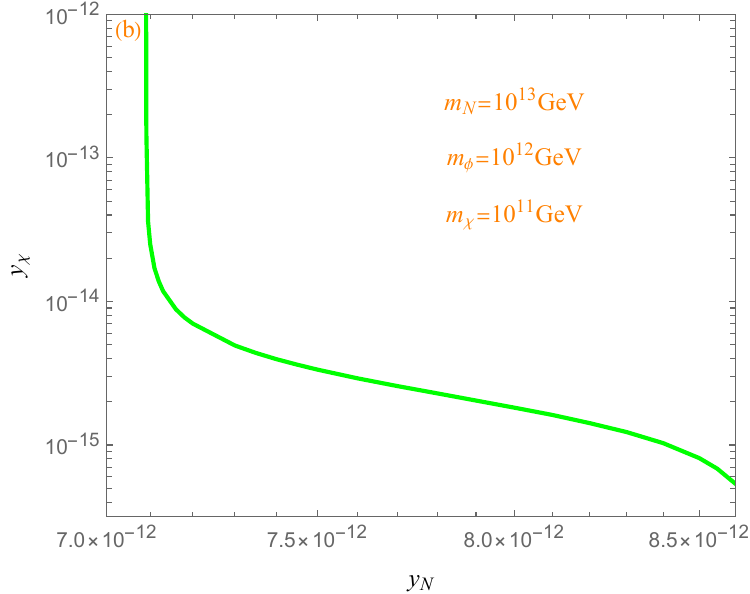}
	\end{center}
	\caption{ The dependence of Yukawa couplings for ordering type B. Panel (a): The required values of the Yukawa coupling $y_N$ as a function of the mass ratio $m_N/m_\chi$. We also fix $y_N=y_\chi$ and $m_N=10^3m_\phi$.  Panel (b): The required values of the Yukawa coupling $y_\chi$ as a function of the Yukawa coupling $y_N$. The mass spectrum is set as $m_N=10^{13}$ GeV, $m_\phi=10^{12}$ GeV, and $m_\chi=10^{11}$ GeV. }
	\label{fig5}
\end{figure}

The dark matter abundance in Equation \eqref{Eqn:RDB} indicates that the Yukawa coupling $y_N$ is determined by the mass ratio $m_N/m_\chi$ for ordering type B. In panel (a) of Figure \ref{fig5}, we show the required value of Yukawa coupling $y_N$ as a function of the mass ratio $m_N/m_\chi$ for correct dark matter relic abundance. It is clear that increasing the mass ratio $m_N/m_\chi$ requires a larger $y_N$.
In panel (b) of Figure \ref{fig5}, the impact of the new Yukawa coupling $y_\chi$ in the $Z_3$ model is illustrated. Typically, a larger new Yukawa coupling $y_\chi$ requires a smaller $y_N$. Because the three-body decay $\phi\to \chi h\nu$ is suppressed by the phase-space, we find that the $\phi\to \chi h \nu$ is the dominant decay mode of dark scalar only when the coupling $y_\chi$ is about two orders of magnitudes smaller than the coupling $y_N$. On the other hand, when $y_\chi\gtrsim y_N/10$, the $\phi\to\chi\chi$ will become the dominant decay mode of dark scalar.

\subsection{Impact of Reheating Temperature}

The results in previous sections \ref{SEC:OTA} and \ref{SEC:OTB} are under the assumption of high reheating temperature $T_\text{RH}\gg m_\phi,m_N$, such that the dark matter production does not depend on $T_\text{RH}$. As we consider the super heavy dark particles being around natural seesaw scale $\sim\mathcal{O}(10^{14})$ GeV, the scenario with reheating temperature smaller than the dark particle or right-handed neutrino mass might happen. In this case, the production of dark matter should be modified according to Equation \eqref{Eqn:RHS} and \eqref{Eqn:RHD}, where the interaction rates could be kinematically suppressed. Therefore, the reheating temperature would have a great impact on the dark matter production.

\begin{figure}
	\begin{center}
		\includegraphics[width=0.45\linewidth]{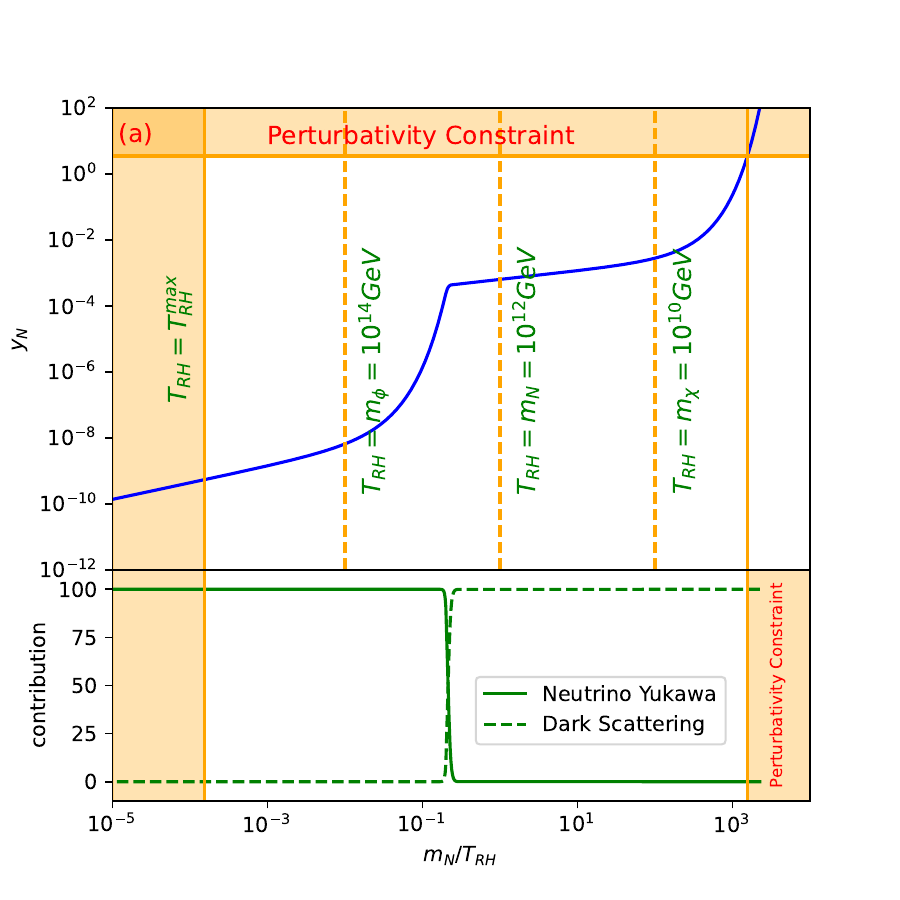}
		\includegraphics[width=0.45\linewidth]{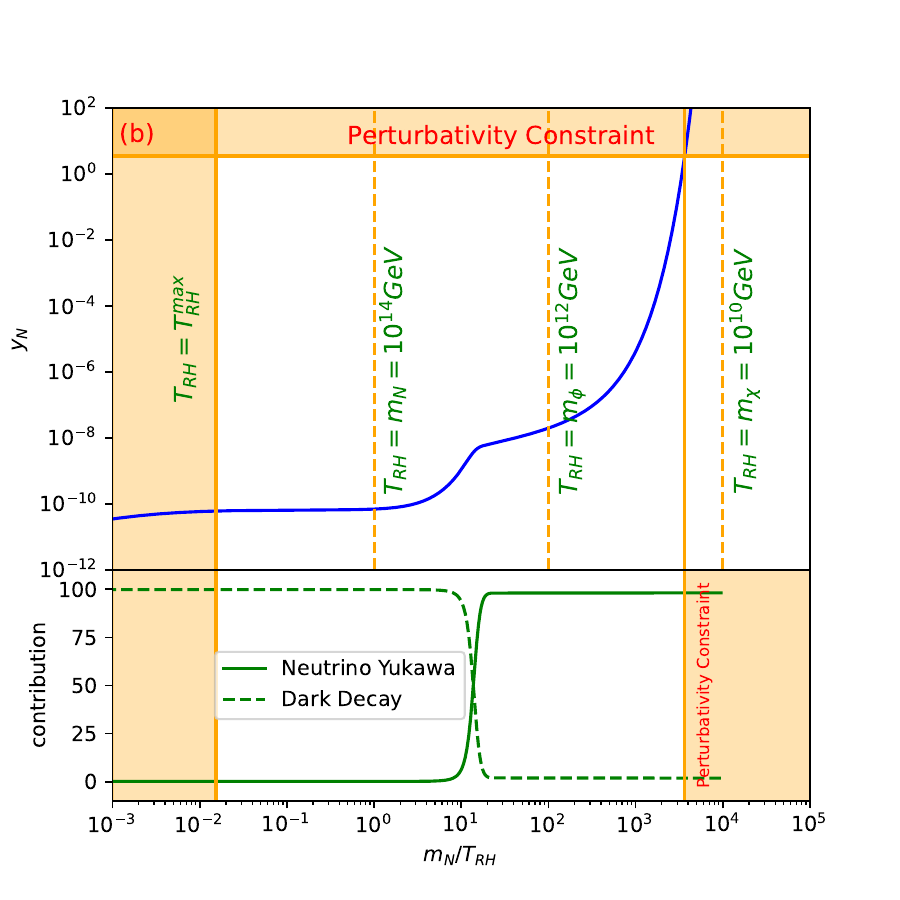}
	\end{center}
	\caption{ The required values of the Yukawa coupling $y_N$ as a function of the reheating temperature for ordering type A in panel (a) and type B in panel (b). In both scenarios, we keep $y_N=y_\chi$. The shaded orange regions are not allowed by the perturbativity constraint $y_N\leq \sqrt{4\pi}$. The lower parts of the figures are the relative contribution of certain processes to the dark matter relic abundance.}
	\label{fig6}
\end{figure}

In Figure \ref{fig6}, we illustrate the dependence of the Yukawa coupling $y_N$ to generate the observed dark matter relic abundance with the reheating temperature $T_\text{RH}$ for both mass ordering. Generally speaking,  when the reheating temperature is much higher than all particles, the value of the Yukawa coupling $y_N$ is the smallest and remains constant. As the equilibrium abundance scales as $Y_i^\text{eq}\propto(m_i/T)^{3/2} e^{- m_i/T}$, the interaction rates in Equation \eqref{Eqn:RHS} and \eqref{Eqn:RHD} are exponentially suppressed  when $T\lesssim m_i$. Therefore, when the reheating temperature decreases to less than the mass of certain particle, the Yukawa coupling $y_N$ begins to increase dramatically. However, there are obvious critical points for both mass ordering, mainly due to the different dependence of the Yukawa coupling $y_N$ and temperature $T_\text{RH}$ for different processes. Very low reheating temperature much smaller than the dark sector is not favored, as the relic density required Yukawa coupling $y_N$ would exceed the perturbation limit $y_N\leq \sqrt{4\pi}$.

For the mass ordering of type A, we choose the spectrum $m_\phi>m_N>m_\chi$ for illustration. In the high reheating temperature limit, the neutrino Yukawa scattering $\phi\chi\to h\nu$ plays the dominant role. Contribution of the scattering $\phi\chi\to h\nu$ is clearly suppressed when $T_\text{RH}<m_\phi$. As we fixed $m_N=10^{12}$ GeV, the corresponding neutrino Yukawa coupling $y_\nu$ is at the order of $\mathcal{O}(10^{-2})$ for the benchmark point in Figure \ref{fig6}~(a). After the critical point, the dark scattering $\chi\chi\to NN$ with $y_N\gtrsim\mathcal{O}(10^{-3})$ becomes the dominant contribution. When $T_\text{RH}$ is smaller than $m_\chi$, the $\chi\chi$ scattering process is also kinematically suppressed, thus the Yukawa coupling $y_N$ quickly becomes larger than $\sqrt{4\pi}$.

The result of the mass ordering type B is shown in panel (b) of Figure \ref{fig6}. For the reheating temperature larger than the right-handed neutrino mass, the direct on-shell dark decay $N\to \phi\chi$ is the dominant contribution. However, the contribution from the on-shell decay is suppressed when $T_\text{RH}<m_N$. With tiny Yukawa coupling $y_N\sim\mathcal{O}(10^{-10})$,  we also find that the dark scattering processes as $NN\to \chi\chi,\phi\phi$ can always be neglected for the benchmark point.   After the critical point $x_N/T_\text{RH}\sim\mathcal{O}(10)$, the contribution from the off-shell scattering $\phi\chi\to h\nu$ becomes important,  where the neutrino Yukawa coupling coupling $y_\nu$ is at the order of $\mathcal{O}(10^{-1})$ with $m_N=10^{14}$ GeV. The contribution from the off-shell scattering  is also suppressed when $T_\text{RH}<m_\phi$.

\begin{figure}
	\begin{center}
		\includegraphics[width=0.45\linewidth]{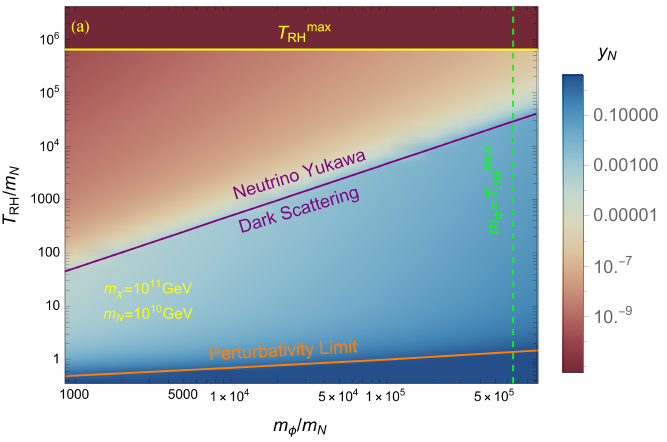}
		\includegraphics[width=0.45\linewidth]{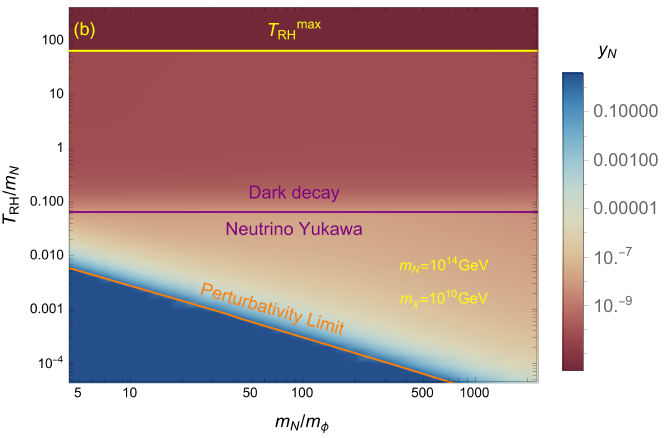}
		\includegraphics[width=0.45\linewidth]{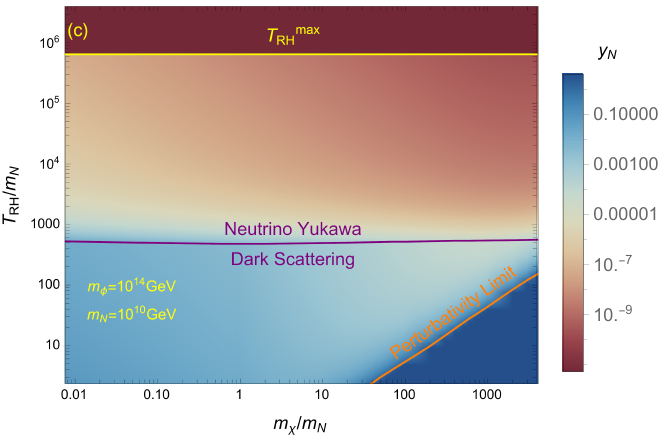}
		\includegraphics[width=0.45\linewidth]{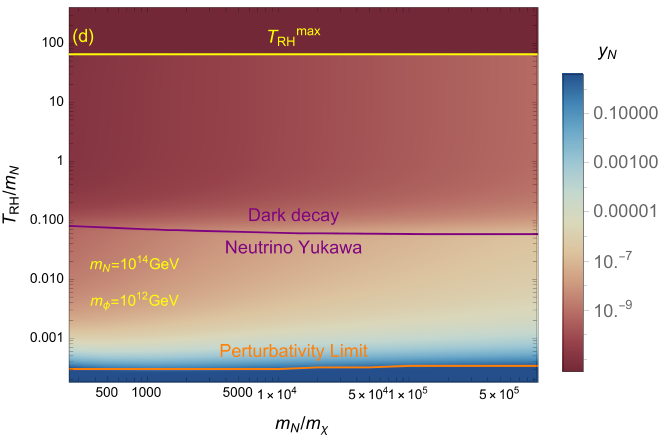}
	\end{center}
	\caption{ The relic density required value of the right-handed neutrino coupling $y_N$ as a function of the dark sector mass and reheating temperature. The left and right panels are the results of mass ordering type A and type B, respectively. The yellow lines are the maximum reheating temperature $T_\text{RH}^\text{max}=6.5\times10^{15}$ GeV. The orange lines are the perturbativity limit $y_N=\sqrt{4\pi}$ . The purple lines correspond to the critical point with equal contribution to relic density of two different processes. We fix $m_\chi=10^{11}$ GeV and $m_N=10^{10}$ GeV in panel (a),  $m_\chi=10^{10}$~GeV and $m_N=10^{14}$~GeV in panel (b), $m_\phi=10^{14}$ GeV and $m_N=10^{10}$ GeV in panel (c), $m_\phi=10^{12}$ GeV and $m_N=10^{14}$~GeV in panel (d) accordingly. }
	\label{fig7}
\end{figure}

In Figure \ref{fig7}, we show the relic density required value of the right-handed neutrino coupling $y_N$ as a function of the dark sector mass and reheating temperature, where we have fixed $y_N=y_\chi$ during the scan. For mass ordering type A $m_\phi>m_N$, the neutrino Yukawa scattering $\phi\chi\to h \nu$ dominants the high reheating temperature region, while the dark sector scattering $\chi\chi\to NN$ is the dominant in the low reheating temperature region. When the mass ratio $m_\phi/m_N$ is larger, the reheating temperature reaching the critical point will be higher. Although the critical point does not clearly depend on the mass ratio $m_\chi/m_N$, increasing the mass ratio $m_\chi/m_N$ will lead to a more stringent perturbativity limit on the reheating temperature. For mass ordering type B $m_N>m_\phi$, the high reheating temperature region is dominant by the on-shell dark decay $N\to \phi\chi$. Below $T_\text{RH}/m_N\sim\mathcal{O}(0.1)$, the dominant contribution is from the off-shell scattering $\phi\chi\to h\nu$. Varying the mass ratio $m_N/m_\phi$  would not alert the critical point significantly. While a smaller mass ratio $m_N/m_\phi$ requires a higher reheating temperature to satisfy the perturbation limit. Meanwhile, the critical point and perturbation limit are not sensitive to the mass ratio $m_N/m_\chi$.

It should be noted that although the correct dark matter relic density is possible even with the reheating temperature $T_\text{RH}$ much smaller than the right-handed neutrino mass for ordering type B $m_N>m_\phi$, the success of canonical thermal leptogenesis requires the reheating temperature to be larger than the right-handed neutrino mass \cite{Chianese:2019epo, Hahn-Woernle:2008tsk}. An alternative pathway to generate the observed baryon asymmetry with low reheating temperature is the non-thermal leptogenesis mechanism \cite{Lazarides:1990huy,Murayama:1992ua}, where the right-handed neutrino is produced from the inflaton decay \cite{Barman:2021ost}.
\section{Conclusion}\label{Sec:CL}

The right-handed neutrino $N$ is introduced in the seesaw model to generate the tiny neutrino mass, which is required to be high scale by the naturally large neutrino Yukawa coupling $y_\nu$ and leptogenesis. Such heavy right-handed neutrino can act as the messenger between the dark sector and the standard model through the freeze-in mechanism. In this paper, we study the seesaw portal to super heavy dark matter with the $Z_3$ symmetry. The dark sector contains one dark scalar $\phi$ and one dark fermion $\chi$, which transforms as $\phi\to e^{i2\pi/3} \phi$  and $\chi\to e^{i2\pi/3}\chi$ under the $Z_3$ symmetry.

One crucial problem in the $Z_2$ symmetric model is that the dark scalar $\phi$ is long-lived when $m_N>m_\phi$, which is excluded by BBN due to the hadronic decay of super heavy dark scalar $\phi\to \chi h \nu$. In the $Z_3$ symmetric model, the additional Yukawa interaction $y_\chi \phi\bar{\chi}^c \chi$ induces the new decay mode $\phi\to\chi\chi$. We show explicitly that even with feeble coupling $y_\chi$, the lifetime of dark scalar $\phi$ via the decay $\phi\to \chi\chi$ is enough to be smaller than the time of BBN. Thus, the mass ordering $m_N>m_\phi$ is also viable in the $Z_3$ symmetric model.

Typically, the super heavy dark matter $\chi$ is generated via the neutrino Yukawa scattering $\phi\chi\to h\nu $ in ordering type A $m_\phi>m_N$ and via the dark decay $N\to \phi\chi$ in ordering type  B $m_N>m_\phi$ when the reheating temperature is high above the seesaw and dark sector. The new decay $\phi\to\chi\chi$ makes the conversion of dark scalar more efficient in the $Z_3$ symmetry, so the produced dark matter relic density is larger than in the $Z_2$ symmetry. For the mass ordering type A, the required Yukawa coupling $y_N$ is closely related to the right-handed neutrino mass $m_N$. For the mass ordering type B, the required Yukawa coupling $y_N$ depends on the mass ratio $m_N/m_\chi$. The new Yukawa interaction $y_\chi \phi \bar{\chi}^c \chi$ has a relatively small impact on the required Yukawa coupling $y_N$. Meanwhile, the reheating temperature $T_\text{RH}$ has a great impact on the Yukawa couplings for correct relic abundance.

\section*{Acknowledgments}
This work is supported by the National Natural Science Foundation of China under Grant No. 11805081, Natural Science Foundation of Shandong Province under Grant No. ZR2022MA056 and ZR2024QA138, the Open Project of Guangxi Key Laboratory of Nuclear Physics and Nuclear Technology under Grant No. NLK2021-07, University of Jinan Disciplinary Cross-Convergence Construction Project 2024 (XKJC-202404).

%%%%%%%%%%%%%%%%%%%%%%%%%%%%%

\end{document}